\documentclass[10pt,a4paper]{article}
\usepackage{graphpap,epsfig,amssymb,graphicx,textcomp,amsmath}
\usepackage{multirow}

\begin{document}

\begin{center}
{\Large {\bf Effects of geometry, boundary condition and dynamical rules on the magnetic relaxation of Ising ferromagnet}}

\end{center}

\vskip 1cm

\begin{center}{\it Ishita Tikader$^1$, \it Olivia Mallick$^2$ and \it Muktish Acharyya$^{3,*}$}\\
{\it Department of Physics, Presidency University,}\\
{\it 86/1 College Street, Kolkata-700073, INDIA}\\
{$^1$E-mail:ishita.rs@presiuniv.ac.in}\\
{$^2$E-mail:olivia.rs@presiuniv.ac.in}\\
{$^3$E-mail:muktish.physics@presiuniv.ac.in}\end{center}

\vskip 1cm

\noindent {\bf Abstract:} We have studied the magnetic relaxation behavior of a two-dimensional Ising
ferromagnet by Monte Carlo simulation. Our primary goal is to investigate the effects of
the system's geometry (area preserving) , boundary conditions, and dynamical rules on the relaxation
behavior. The Glauber and Metropolis dynamical rules have been employed. The systems with periodic and
open boundary conditions are studied. The major findings are the exponential relaxation and the dependence of relaxation time ($\tau$) on the aspect ratio $R$ (length over breadth having fixed area). 
A power law dependence ($\tau \sim R^{-s}$) has been observed for larger values of aspect ratio ($R$). The exponent ($s$) has
been found to depend linearly ($s=aT+b$) on the system's temperature ($T$). The transient behaviours of the spin-flip
density have been investigated for both surface and bulk/core. The size dependencies of saturated spin-flip density
significantly differ for the surface and the bulk/core. Both the saturated bulk/core 
and  saturated surface spin-flip density was found to follow
the logarithmic dependence $f_d = a + b~log(L)$ with the system size. The faster relaxation was
 observed for open boundary condition with any kind (Metropolis/Glauber) of dynamical rule. 
 Similarly, Metropolis algorithm yields faster relaxation for any 
 kind (open/periodic) of boundary condition.

\vskip 3cm


\vskip 2cm

\noindent $^*$ Corresponding author
\newpage

\noindent {\Large {\bf {1. Introduction:}}}

\vskip 1cm

The cooperatively interacting thermodynamic systems show the relaxation behaviour. If the system is perturbed by an external
agency, it takes some time to reach the original equilibrium 
state, after the removal of the perturbation. The ferromagnetic
samples,  if perturbed
by a strong magnetic field, also relax towards their equilibrium state, after the removal of the magnetic field. The study of this
relaxation behaviour of the ferromagnets became an interesting field of research in the last few decades.

The Ising model is the prototype of ferromagnets. The relaxation behaviour of Ising ferromagnet\cite{richards} is widely studied in recent past.
Although the Ising ferromagnet does not have any intrinsic dynamics, the mean-field dynamical equation was developed by Suzuki
and Kubo \cite{kubo} for Glauber kinetic Ising ferromagnet. This
mean-field dynamical equation provides the model to study the dynamical phenomena, e.g. relaxation behaviour. 

A series of investigations of the relaxation behaviours of Ising ferromagnet may be mentioned as follows: The decay of spin-spin correlation in the low temperature regime in the one dimensional Ising ferromagnet has been studied\cite{prados} using Glauber dynamics. The large scale critical dynamics have been studied
\cite{lin} to
investigate the linear relaxation in two dimensional Ising ferromagnet and estimated the dynamic exponent. The relaxation
of energy in two dimensional Ising model has been investigated\cite{mori} by Monte Carlo simulation and estimated the critical exponents (above critical temperature) of linear and nonlinear relaxations of the internal energy. The linear and nonlinear relaxation behaviour in kinetic Ising model has been
analyzed\cite{collins} by Pade approximation. The tricritical 
relaxation was studied\cite{landau} in Glauber kinetic three dimensional Ising model and found that below the tricritical temperature the
relaxation is fairly exponential in nature. A slow relaxation
has been reported recently \cite{jeong} in Ising small-world networks with strong long-range interactions. The critical nonequilibrium relaxation in Ising model has been studied\cite{tomita} by cluster-flip algorithm. The field and size dependence of the relaxation behaviour of the metastable states of the two dimensional Ising model has been studied\cite{miyashita}. The
slow relaxation has been reported \cite{satya} in a constrained Ising
spin chain. The results obtained from the series-expansion and Monte Carlo methods are compared \cite{wang} in the non-equilibrium relaxation of Ising model. The relaxation of Glauber kinetic Ising model has been
investigated\cite{binder2} via the theory of the dynamics and clusters near the
critical point. The effects of random impurities on the relaxation of Ising ferromagnets have also been investigated\cite{oerding}. The statistical distribution of relaxation time in Ising ferromagnet has been studied\cite{melin} and found the logarithm of the relaxation time is normally distributed in the strong field regime. The relaxation in Ising model has been studied (decay of Hamming distance by Monte Carlo simulations) and observed\cite{stauffer2} a simple exponential decay in three and four dimensions below and above the Curie temperature. In two dimensions, a stretched exponential decay is observed only below the critical temperature. The research on the relaxation behaviour of Ising ferromagnet is not only limited by
spin-${1 \over 2}$. The magnetic relaxation in spin-1 Ising ferromagnet has also been studied\cite{erdem} near the second-order phase
transition point by solving the kinetic equation obtained from molecular field theory. They have obtained the magnetic dispersion and calculated the absorption factor near the second
order phase transition point. The slow relaxation has been observed \cite{nowak} in diluted antiferromagnets by Monte Carlo simulation.

For completeness, let us mention here some of the experimental studies on the relaxation of magnetic samples. The surface magnetic relaxation
was studied\cite{godfrin} in He$^3$ experimentally by NMR technique in confined geometry. The magnetic relaxation has been 
experimentally investigated\cite{zazo} in superconducting 
${\rm YBa_2Cu_3O_7}$ sample. A Carbon nanotube based magnetic nanocomposite sample was chosen to study\cite{rosa} the magnetic relaxation. The magnetic relaxation behaviours are studied experimentally in single-ion magnets\cite{chiesa} and single molecule magnets\cite{lei}. The type of relaxation behaviour was also investigated\cite{suran} experimentally to find the non-Arrhenius (does not obey Neel-Arrhenius law) magnetic relaxation. The highly anisotropic ferromagnetic ultrathin films
have been used to study\cite{dunlavy} the relaxation behaviour (mainly the critical slowing down) in two dimensional
Ising ferromagnets.

In reviewing the above-mentioned studies on the relaxation phenomena
in Ising ferromagnet, we have not found any study regarding the roles of boundary 
conditions, the dynamical rules (Glauber/Metropolis) and the geometrical structures in the relaxation process of Ising ferromagnet. How will the relaxation behaviour get affected by these factors ? In this article, we have addressed these questions and studied the
relaxation of Ising ferromagnet subject to different boundary 
conditions,  geometrical structures and 
dynamical rules. We have organised the manuscript as follows: After the introduction in section-1, the ferromagnetic Ising model and the Monte Carlo simulation method are discussed in section-2. The simulational results are reported in section-3 and
the paper ends with concluding remarks in section-4.

\vskip 1cm

\vskip 1cm

\noindent {\Large {\bf 2. The Ising Model and the Monte Carlo Simulation scheme:}}
\vskip 0.5cm 

We consider a two-dimensional Ising ferromagnetic system with uniform nearest-neighbour interaction described by the following Hamiltonian (in the absence of any external
magnetic field),
\begin{equation}
	H=-J\:\sum_{<i,j>} S_i S_j 
	\label{eqn:19}
\end{equation}	
where Ising spin variables $S_i$ can access two discrete values, i.e., +1 and -1 only. $J$ is the uniform ferromagnetic ($J>0$) interaction strength between the nearest neighbour spin pairs.

The system is considered to be a two-dimensional rectangular lattice, having a total number of spins $N=L_x\times L_y$. Here, $L_x$  and $L_y$ represent the length along X-axis and Y-axis respectively and for square lattice $L_x=L_y$. The aspect ratio is
$R={{L_x} \over {L_y}}$.
Here, we have studied the relaxation behaviours of two dimensional Ising ferromagnet for different aspect ratios ($R$) but having constant area ($L_x \times L_y=4096$). Since we are interested to study the role of boundary conditions applied to the Ising system, we have considered  both periodic (PBC) and open (OBC) boundary conditions.

Now let us briefly discuss the Monte Carlo simulation method
\cite{binder}we have employed here. The system is initially kept in a perfectly ordered state, where all spins are pointing up ie $S_i = +1$. This specific configuration may be imagined as the equilibrium state of the system in the presence of very high magnetic field at any finite (nonzero) temperature. The system is in contact with a heat bath at a constant temperature ($T$). Here, the temperature is measured in the unit of $J/k_B$. The target spin (to be flipped or not) has been selected at random. Now the fate of the target spin is determined by the Metropolis single spin flip algorithm or Glauber protocol. Here, we wanted to focus on both kinds of  dynamics. Let us denote the randomly selected target spin by $S_i$. The change in energy due to 
trial spin flip is $\Delta E$ (in units of $J$) is determined by the Hamiltonian (equation-\ref{eqn:19}) mentioned above. The target spin will flip  with transition probability prescribed by Metropolis \cite{metropolis} or Glauber\cite{glauber} algorithm.

Metropolis transition probability is given by, 
\begin{equation}\label{eqn:20}
    P(S_i \rightarrow  -S_i) = Min[1, \exp (-\frac{\Delta E}{k_BT})]
\end{equation}

Glauber transition probability is,  
\begin{equation}\label{eqn:21}
    P(S_i \rightarrow  -S_i) = \frac{\exp (-\frac{\Delta E}{k_BT})}{1+ \exp (-\frac{\Delta E}{k_BT})}
\end{equation}
where $k_B$ is the Boltzmann constant and T is the temperature of the system. We have assumed $k_B=1$ and $J=1$ for simplicity. 

If a randomly chosen number (say r), uniformly distributed in the range [0,1] is less than or equal to spin-flip probability $ P(S_i \rightarrow  -S_i)$ i.e., $r\leq P$, only then  the trial spin would flip (i.e., $S_i \to -S_i$). This protocol is repeated for the total $N= L_x \times L_y$ number of  randomly updated spins. $L_x \times L_y$ number of such random updates is termed as one Monte Carlo Step per Spin or MCSS which defines the unit of time in this simulation study.\\
At each Monte Carlo step average magnetisation of the system is represented by,
\begin{equation}\label{eqn:22}
   m'(t)= \frac{1}{N}\sum_{i=1}^{N} S{_{i}}. 
\end{equation}
The average instantaneous magnetization $m(t)$ has been calculated over 50000 number of different samples.
\vspace{1cm}

\noindent {\Large{\bf 3. Simulational results:}}
\vskip 0.5cm
\noindent {\bf A. Relaxation of magnetisation:}
\vskip 0.5cm

Initially, we have chosen a square lattice ($L_x=L_y$) of size $L{_{x}}\times L{_{y}}$ and considered that all spins are in the up state; i.e., $S{_{i}}=+1$, for all $i$. This situation can be imagined that, a strong magnetic field is applied to the system externally along the positive $z$ direction. All of a sudden we switched off the magnetic field. Now in absence of an external field, the system is ready to relax to its equilibrium state at a fixed temperature $T>T{_{c}}$ (where $T_c$ is the critical temperature). We have thoroughly observed the relaxation of the two dimensional Ising system. In this paper, we are concerned about the role of dynamical algorithms on the transient behaviour of the Ising ferromagnetic system. In particular, our goal is to explore a comparative study between Metropolis algorithm and Glauber dynamics, in the context of magnetic relaxation. We have also studied the effects of 
boundary conditions (open or periodic) on  the magnetic relaxation of the system. We have tried to address the most interesting question. Does the system relax in the same manner if we change the geometry of the lattice keeping the area same? How does this relaxation time change if we deform a square lattice into a quasi-one dimensional strip (keeping the area constant) ? 

\begin{figure}[h!]
\begin{center}

\includegraphics[angle=0,width=0.45\textwidth]{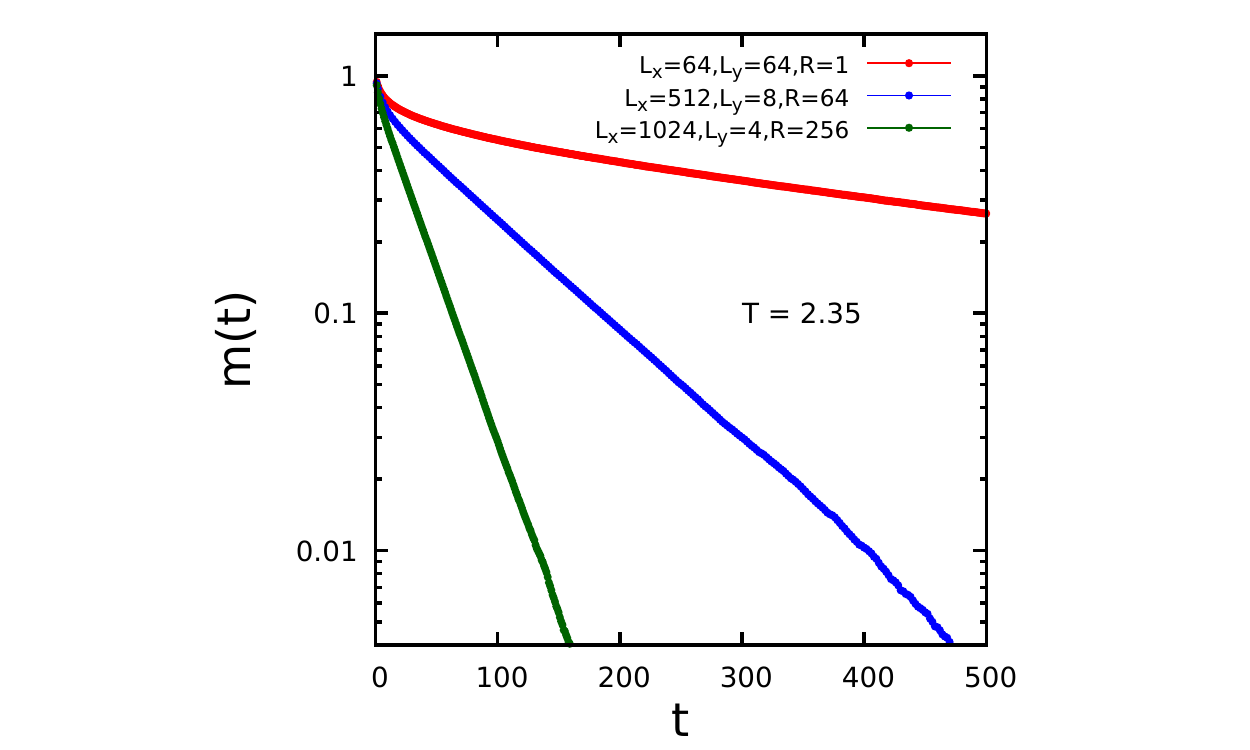}(a)
\includegraphics[angle=0,width=0.45\textwidth]{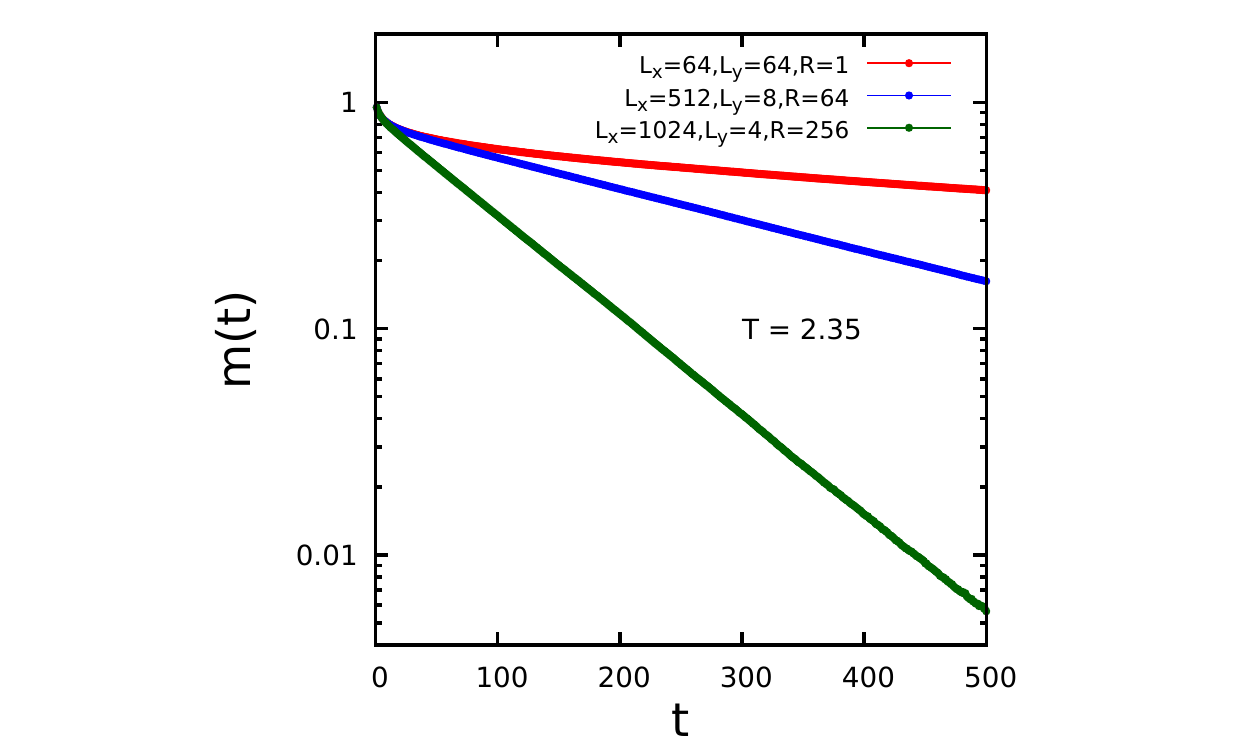}(b)
\\
\includegraphics[angle=0,width=0.45\textwidth]{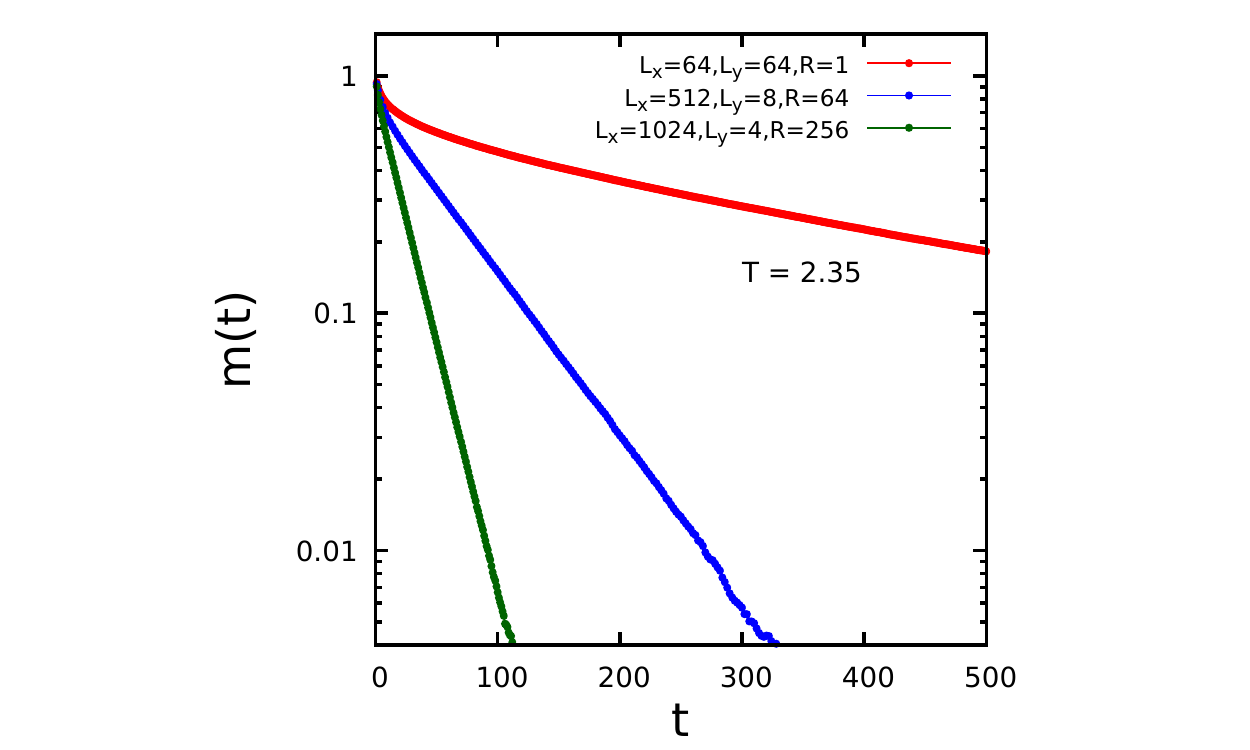}(c)
\includegraphics[angle=0,width=0.45\textwidth]{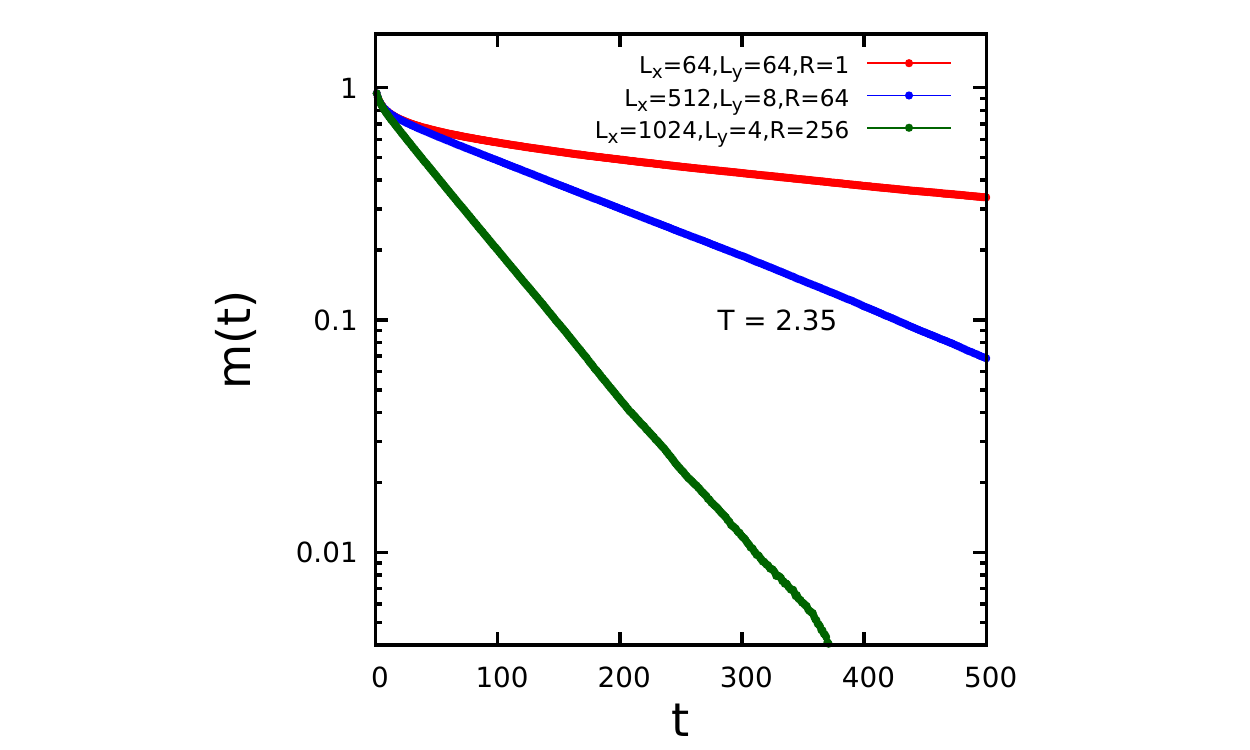}(d)

\caption{The study of the relaxation behaviour for different geometry with the fixed number of spins or fixed area ($L_x \times L_y$=4096 here) and temperature $T=2.35$. We have ploted the data in semi-logarithmic scale. (a) open boundary condition using Glauber dynamics, (b) periodic boundary condition using Glauber dynamics, (c) open boundary condition using Metropolis dynamics and
(d) periodic boundary condition using Metropolis dynamics.} 
\label{same-area}
\end{center}
\end{figure}

 We have observed the temporal evolution of the sample averaged instantaneous magnetisation $m(t)$. Obviously, the average magnetisation has been observed to decay with time. We have
 kept the temperature $T=2.35$. Since, the system is in paramagnetic phase
 ($T_c=2.269...$ is the Onsager value), the magnetisation eventually was found to vanish exponentially, i.e., $m(t) \sim e^{(-t/{\tau})}$ (straight line in semilogarithmic plot). The exponential relaxation was theoretically predicted in the Glauber kinetic Ising ferromagnet\cite{kubo} with meanfield approximation. We have shown these results in Fig-\ref{same-area} for different sizes 
 ($L_x$ and $L_y$ but for $L_x \times L_y=4096$) of the system for
 different boundary conditions. We have studied the relaxation behaviours for different
 dynamical rules (Glauber/Metropolis). For same boundary conditions, the relaxation
 is observed to be faster in the case of Metropolis rule. It is evident from 
 Fig-\ref{same-area}(a) (Glauber) and Fig-\ref{same-area}(c) (Metropolis). The reason is quite clear as the Metropolis probability
can assure the spin flip (for negative change in energy due to flip) which is not provided by Glauber protocol. On the other hand, for the same kind of dynamics, the system with open boundary conditions relaxes faster than that for periodic boundary conditions. It would be clear, if one 
 compares the results in the Fig-\ref{same-area}(c) and Fig-\ref{same-area}(d).

\begin{figure}[h!]
\begin{center}

\includegraphics[angle=0,width=0.45\textwidth]{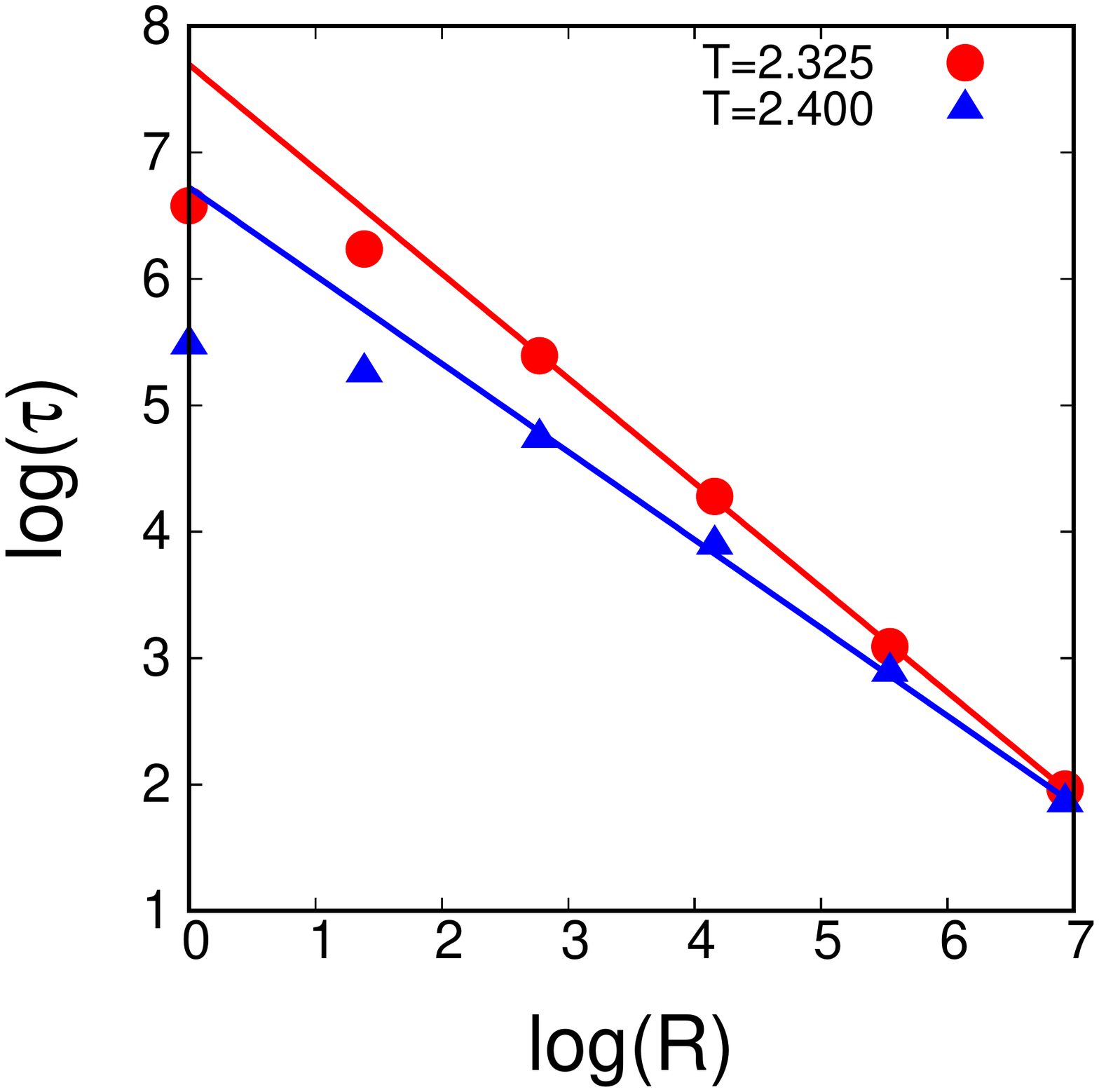}(a)
\includegraphics[angle=0,width=0.45\textwidth]{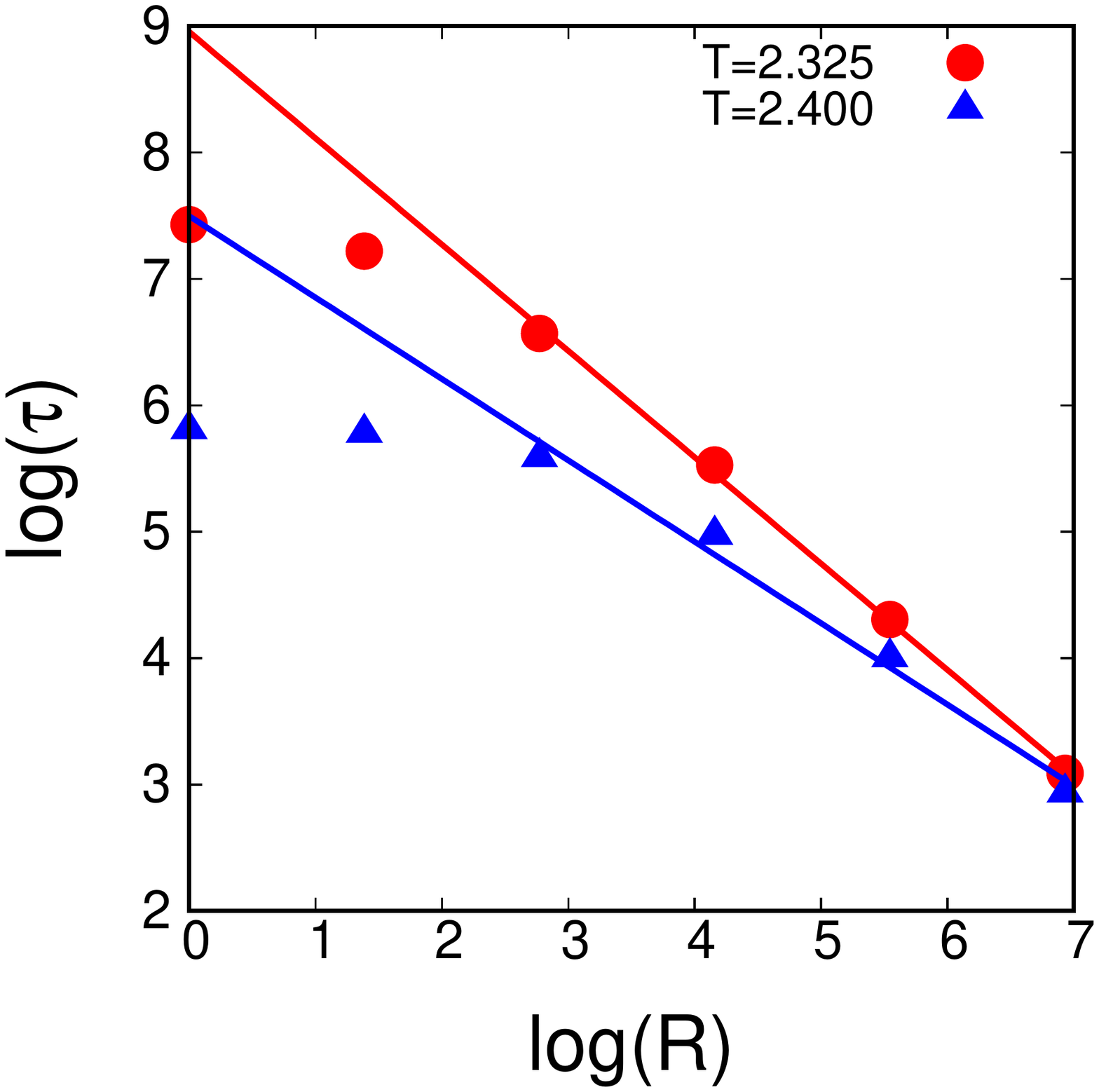}(b)
\\

\includegraphics[angle=0,width=0.45\textwidth]{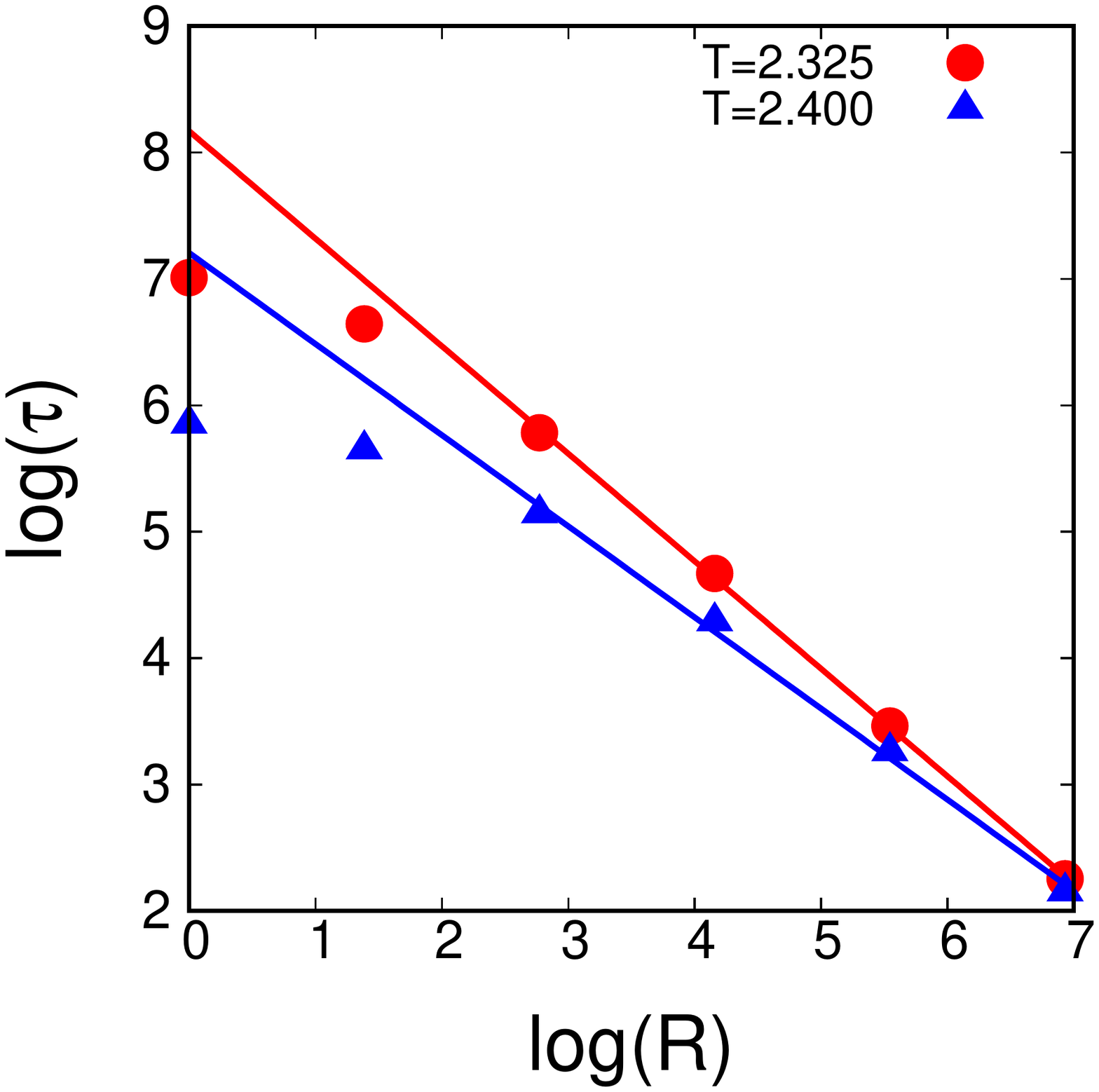}(c)
\includegraphics[angle=0,width=0.45\textwidth]{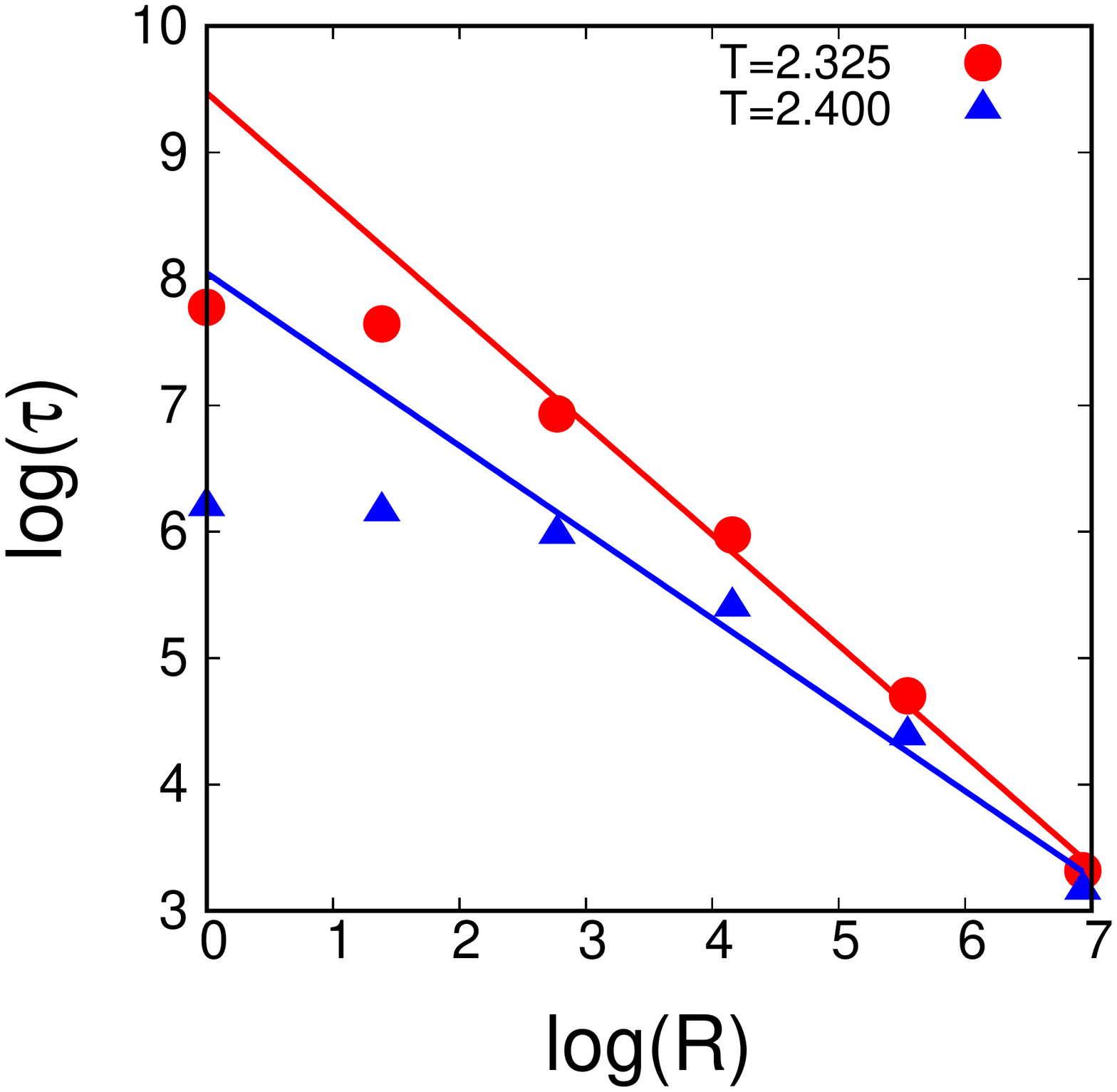}(d)

\caption{The logarithm of Relaxation time ($log(\tau)$) is plotted against 
the logarithm of the aspect ratio ($log(R=L_x/L_y))$ for two different temperatures in the paramagnetic phase, $T=2.325$ (Red circle) and $T=2.400$ (Blue triangle). Both are 
fitted to the straight line , $log(\tau)=-s\times log(R)+r$.
(a) {\bf For open boundary conditions and Metropolis dynamics:} At $T=2.325$;	$s=0.827\pm 0.007, r=7.69\pm 0.04$ (Red). Here DOF=2 and ${\chi }^2 = 0.001$. And at $T=2.400$;	$s=0.696\pm 0.022, r=6.72\pm 0.11 $ (Blue). Here DOF= 2 and ${\chi }^2 = 0.010$. 
(b){\bf For periodic boundary conditions and Metropolis dynamics:} At $T=2.325$;	$s=0.841\pm 0.022, r=8.95\pm 0.11$ (Red). Here DOF=2 and ${\chi }^2 = 0.009$. And at $T=2.400$;	$s=0.644\pm 0.054, r=7.50\pm 0.27 $ (Blue). Here DOF=2 and ${\chi }^2 = 0.056$.
(c){\bf For open boundary conditions and Glauber dynamics:} At $T=2.325$; $s=0.850\pm 0.012, r=8.17\pm 0.06$ (Red). Here DOF=2 and ${\chi }^2 = 0.003$. And at $T=2.400$;	$s=0.721\pm 0.029, r=7.21\pm 0.15$ (Blue). Here DOF=2 and ${\chi }^2 = 0.017$. 
(d){\bf For periodic boundary conditions and Glauber dynamics:} At $T=2.325$;	$s=0.874\pm 0.050, r=9.47\pm 0.25$ (Red). Here DOF=2 and ${\chi }^2 = 0.047$. And at $T=2.400$;	$s=0.683\pm 0.075, r=8.05\pm 0.38 $ (Blue). Here DOF=2 and ${\chi }^2 = 0.107$.}	
\label{tau-R}
\end{center}
\end{figure}

How the relaxation behaviour will depend on the geometry of the lattice? Starting
from a square lattice, if the system is deformed (area preserving) to a rectangular shape, how does the relaxation time ($\tau$) get affected
by such area preserving deformation ? This requires to measure the relaxation time
for various values of $L_x$ and $L_y$ (with $L_x \times L_y$ fixed). In this study, 
 we  gradually changed the geometry of the lattice and deformed the square lattice into a rectangular one (keeping $L_x \times L_y$ fixed). We have measured magnetisation $m(t)$ of Ising ferromagnetic system for different shapes of the lattice by changing the length along X-axis, $L{_{x}}=64$, 128, 256, 512, 1024 and 2048 and changing corresponding values of $L_y$ to keep $L_x \times L_y$=4096. We have defined the aspect ratio
 of the lattice geometry by $R={{L_x}\over{L_y}}$ and studied the logarithm of the relaxation time
 (log($\tau$)) as a function of log($R$) for different boundary conditions
 (open and periodic) and for different
 dynamical rules (Glauber and Metropolis). We have presented our result in a compact form in Fig-\ref{tau-R}.
 The logarithm of the relaxation time (log($\tau$)) is plotted against the logarithm of the aspect ratio (log($R$)). The results obtained by using Metropolis dynamics are shown in
 Fig-\ref{tau-R}(a) and Fig-\ref{tau-R}(b) for open and periodic boundary conditions
 applied to the system respectively. Similarly, the results for open and periodic
 boundary conditions are shown respectively in Fig-\ref{tau-R}(c) and  Fig-\ref{tau-R}(d)  obtained by using Glauber rule. All the results are obtained for
 two different temperatures ($T=2.325$ and $T=2.400$). We have observed that for larger values of $R$, the relaxation time $\tau$ shows a strong
 power law dependence on it ($R$).  This 
 prompted us to propose a relation like, $\tau \sim R^{-s}$. The exponent,
 $s$ of this power law variation, is temperature dependent ($s(T)$). As the temperature
 increases the value of the exponent $s$ decreases for any particular type of
  boundary condition
 and the dynamical rule. For any fixed temperature and boundary condition, the
 value of the exponent $s$ is slightly larger for Glauber dynamics.  The exponent $s$ is found to depend
 on the temperature ($T$).
 
\begin{figure}[h!]
\begin{center}

\includegraphics[angle=0,width=0.45\textwidth]{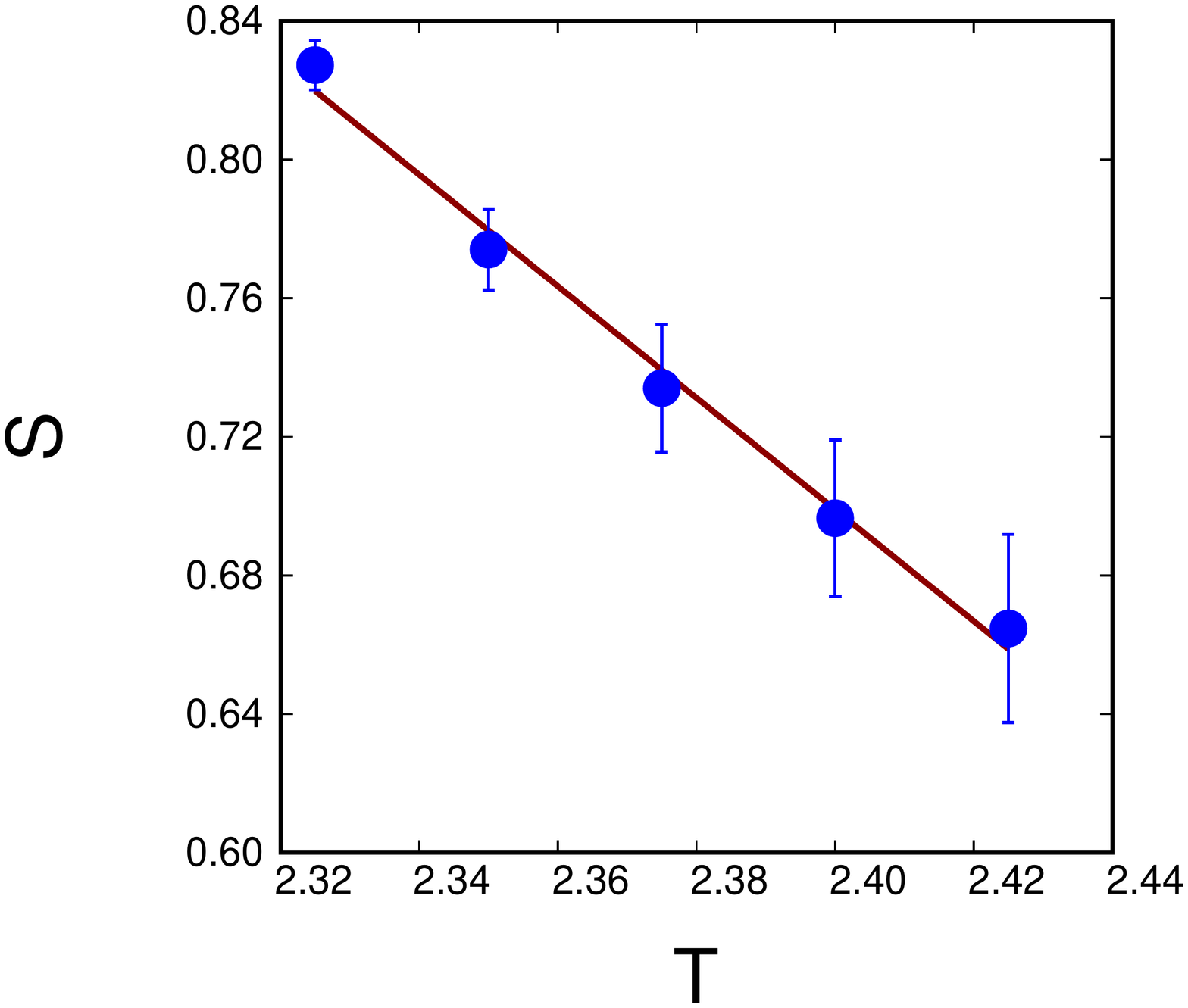}(a)
\includegraphics[angle=0,width=0.45\textwidth]{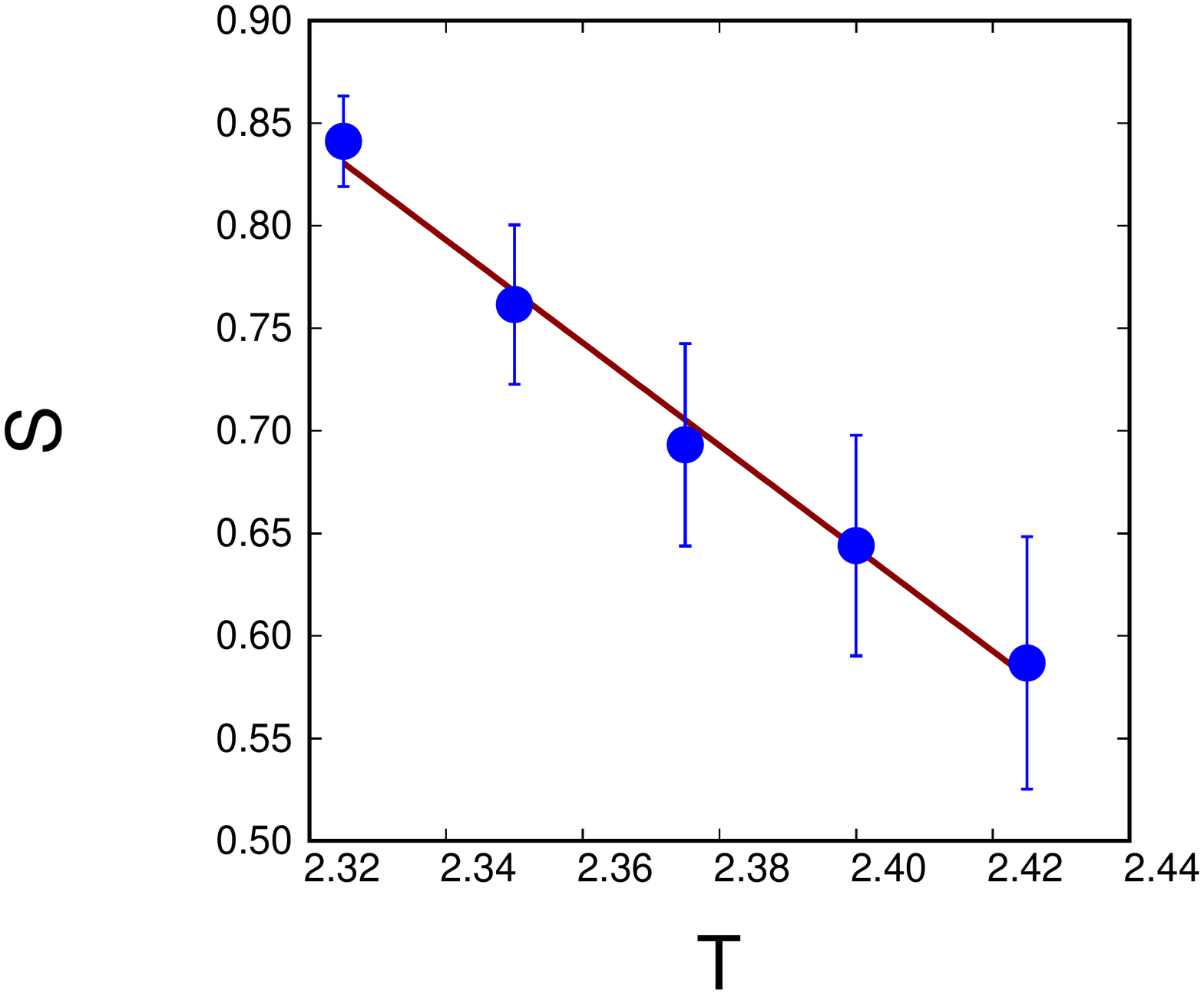}(b)
\\
\includegraphics[angle=0,width=0.45\textwidth]{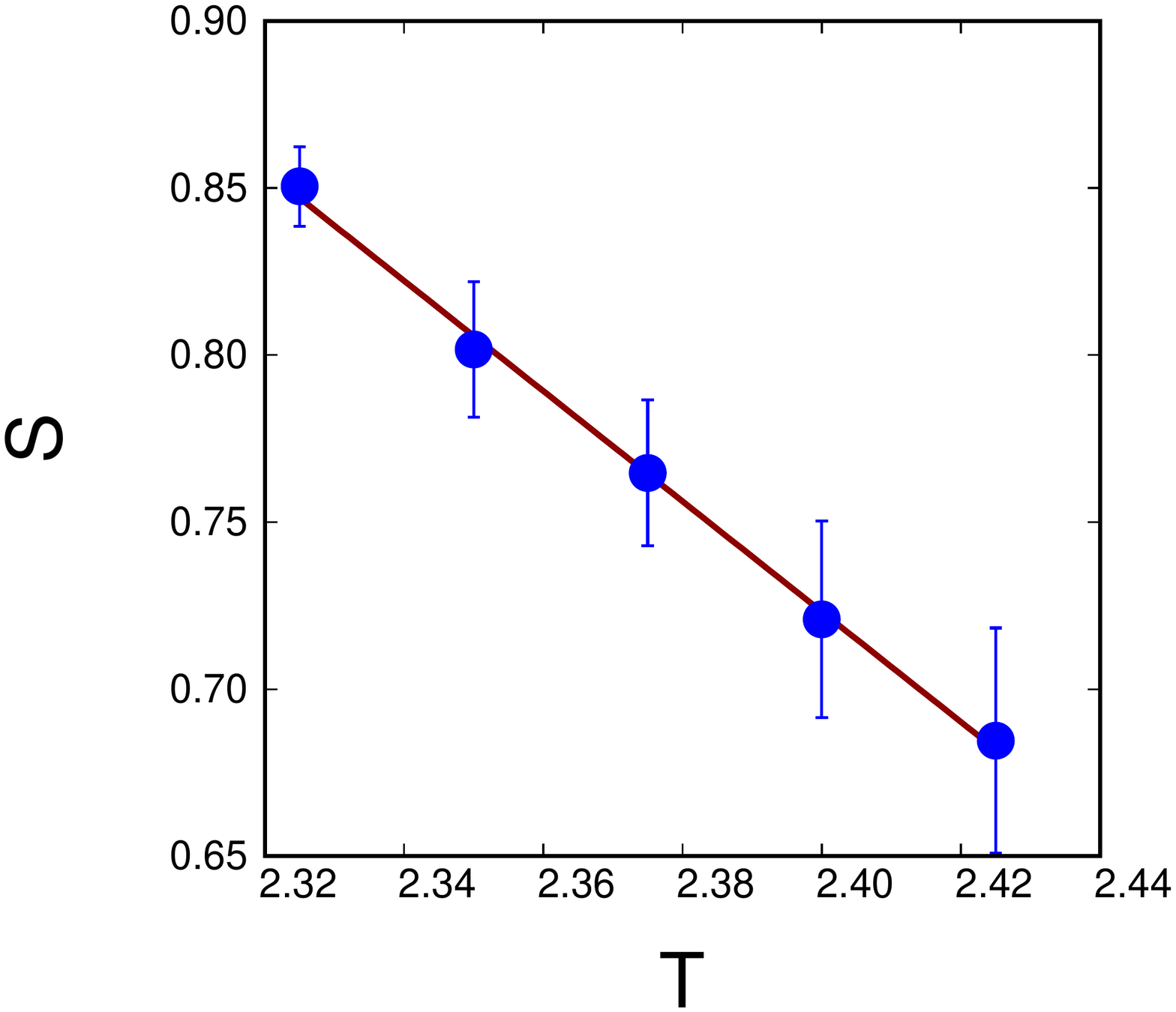}(c)
\includegraphics[angle=0,width=0.45\textwidth]{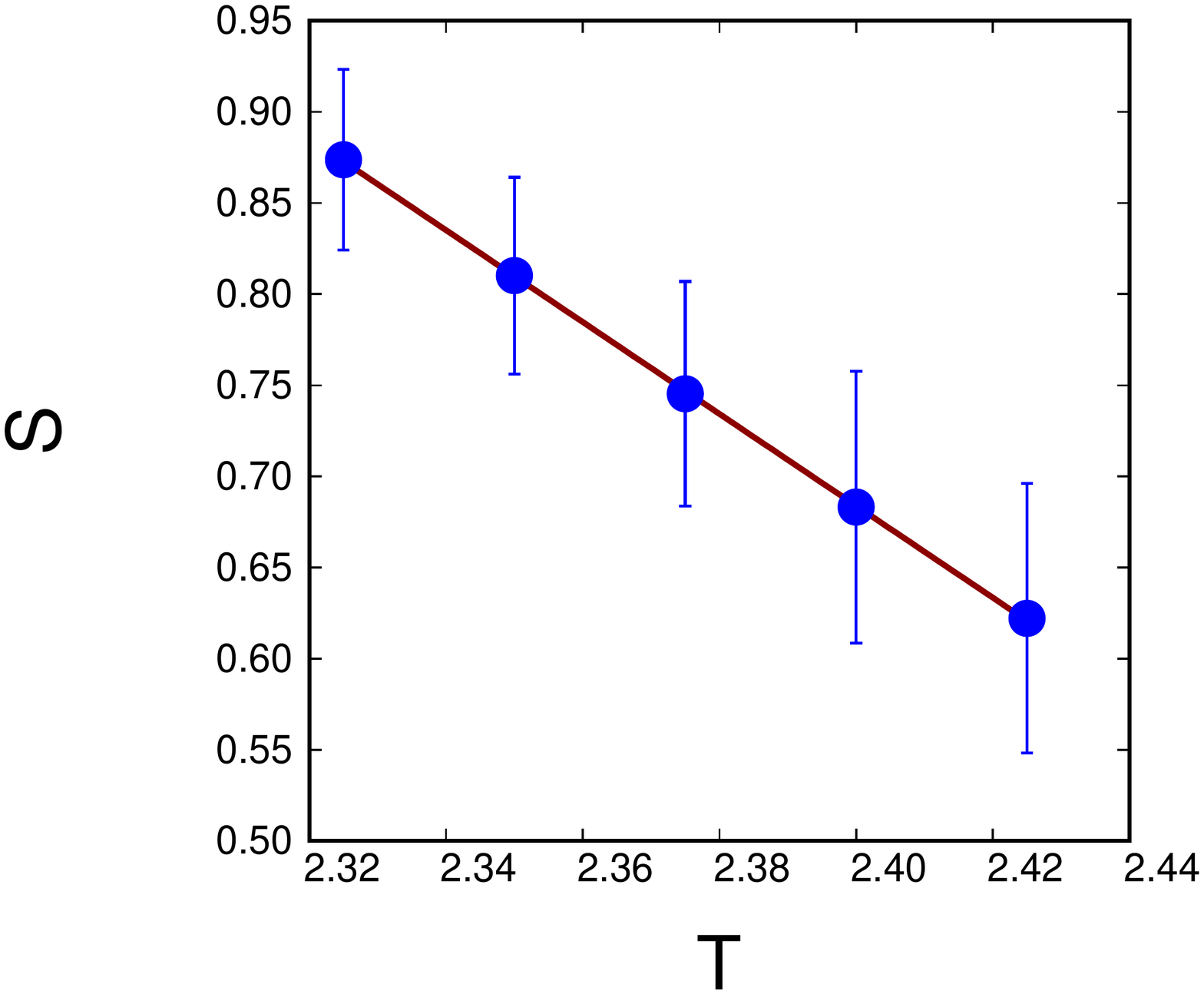}(d)

\caption{The exponent $s$ is plotted against the temperature $T$ (in the paramagnetic phase). The data are fitted to a straight line, $s=aT+b$. 
(a) {\bf For the open boundary conditions and the Metropolis algorithm:} $a= -1.61\pm 0.09, b= 4.56\pm 0.22$. Here DOF= 3 and ${\chi }^2 = 0.0001$. 
(b) {\bf For the periodic boundary conditions and the Metropolis algorithm:} here $a= -2.51\pm 0.14, b= 6.66\pm 0.32$. Here DOF= 3 and ${\chi }^2 = 0.0003$.
(c) {\bf For the open boundary conditions and the Glauber algorithm:} here $a= -1.65\pm 0.05, b= 4.68\pm 0.11$. Here DOF= 3 and ${\chi }^2 = 4.1\times 10^{-5}$. 
(d){\bf For the periodic boundary conditions and the Glauber algorithm:} here $a= -2.52\pm 0.02, b= 6.73\pm 0.04$. Here DOF= 3 and ${\chi }^2 = 5.2\times 10^{-6}$.} 
\label{exponent-T}
\end{center}
\end{figure}

 What will be the precise thermal ($T$) variation of the exponent $s$ ? To get the
 answer to this question, we have estimated the exponent $s$ ($\tau \sim R^{-s}$) for
 a range of temperatures (in the paramagnetic region) and shown the
 results in Fig-\ref{exponent-T}. The results obtained by using Metropolis dynamics
 have been shown in Fig-\ref{exponent-T}(a) and in Fig-\ref{exponent-T}(b) for open 
 and periodic boundary conditions respectively. Similarly, the results obtained by
 Glauber rule are shown in Fig-\ref{exponent-T}(c) and in Fig-\ref{exponent-T}(d) 
 for open and periodic boundary conditions respectively. Our results shows that
 $s=aT+b$ (linear variation with temperature). The results obtained by linear
 regression and the fit-statistics are given in Table-\ref{Table1}. Qualitatively, similar behaviour has been observed for both Metropolis and Glauber dynamics.
 However, the rate of linear fall ($-{{ds} \over {dT}}$) of the exponent $s$ with
 respect to the temperature is 
 significantly higher in the case of periodic boundary condition.

   Let us now systematically present our results for different
dynamics and different boundary conditions in a tabular form (Table-\ref{Table1}).

\begin{table}[h!]
\begin{center} {\bf{Table 1}} \\
 \vspace{0.2cm}   
\begin{tabular}{|p{2.5cm}|p{3.2cm}|p{2cm}|p{2cm}|p{2cm}|p{1cm}|}
  \hline
  Dynamical rule & Boundary condition  & a & b & ${\chi }^2$  & DoF \\ [1ex]
  \hline 
   \multirow{2}{*}{Metropolis} & Open & $-1.61\pm 0.09$ & $4.56\pm 0.22$ & 0.0001 &  3 \\ [1ex] \cline{2-6}
   & Periodic & $-2.51\pm 0.14$ & $6.66\pm 0.32$ & 0.0003 & 3 \\ [1ex]
   \hline
   \multirow{2}{*}{Glauber} & Open & $-1.65\pm 0.05$ & $4.68\pm 0.11$ &  $4.1\times 10^{-5} $&  3 \\ [1ex] \cline{2-6} 
   & Periodic & $-2.52\pm 0.02$ & $6.73\pm 0.04 $ & $5.2\times 10^{-6}$ & 3 \\ [1ex]
   \hline
  \end{tabular}
 \caption{Fitting parameters for the exponent $s$ versus temperature $T$ plot.}
 \label{Table1}
\end{center}
\end{table}

\vskip 0.5cm
\noindent {\bf B. Temporal Growth of Spin flip density:}
\vskip 0.5cm

Since, the decay of the metastable (or unstable) state of the cooperatively interacting
spin system is microscopically governed by spin-flip, the average number of spin-flip
per site (spin-flip density), would be a crucial measure of relaxation dynamics. We 
have also studied the spin-flip density as function of time. As the system relaxes,
the accompanying spin-flip density grows in time. In the paramagnetic phase, the spin-flip density eventually saturates with some fixed values after relaxation. 
This saturated spin-flip density is the function of the temperature and  system sizes. Precisely, it depends on the surface to volume ratio.
 
We consider a  two dimensional square lattice. The system consists of $N=L^2$ number of spins where the surface contains $N_s=(4L-4)$ number of spins. The inner part of the lattice, defined as the core, contains $N_c=(L-2)^2$ number of spins. Each spin inside the core has four nearest neighbours; each spin on the edge of the lattice has three nearest neighbours and each of the four spins at corners has two nearest neighbours only. Initially, the system is in a perfectly ordered state with all spins pointing up ($S^z_i=+1$ for all $i$). In order to inspect the free surfaces, the so-called open boundary condition is incorporated here.  The spin-flip probability is governed by Metropolis\cite{metropolis} and Glauber \cite{glauber} dynamics exclusively. The spin-flip density $f_d$ is defined  in the following way :

\begin{equation}
    f_d(t)=\frac{\text{\rm Number~of~spin-flips(surface/core)~at~t-th~MCSS}}{\text{\rm total~number~of ~sites~on~surface/core}}
 \end{equation} 

\begin{figure}[h!]
\begin{center}

\includegraphics[angle=0,width=0.4\textwidth]{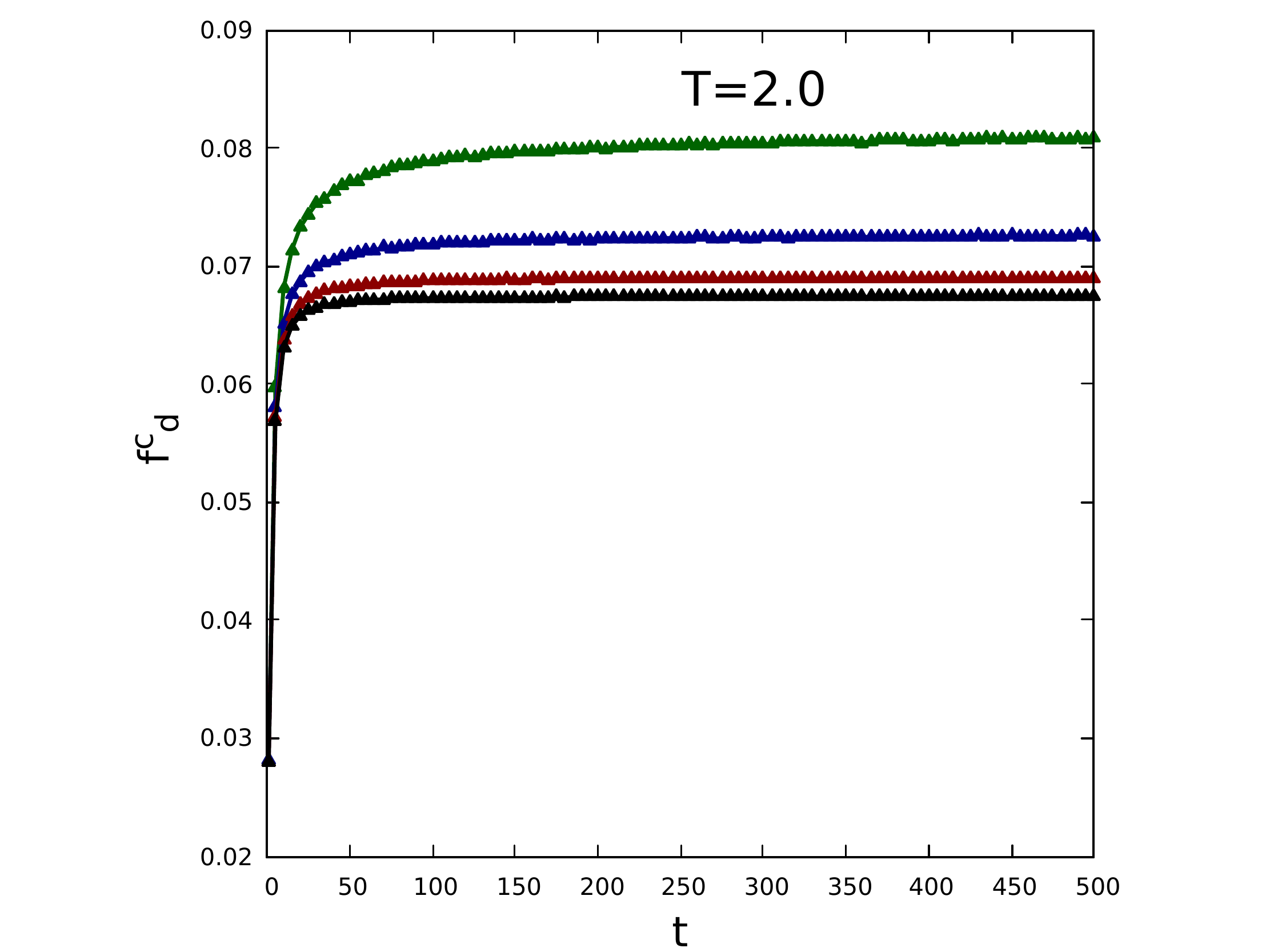}
(a)
\includegraphics[angle=0,width=0.4\textwidth]{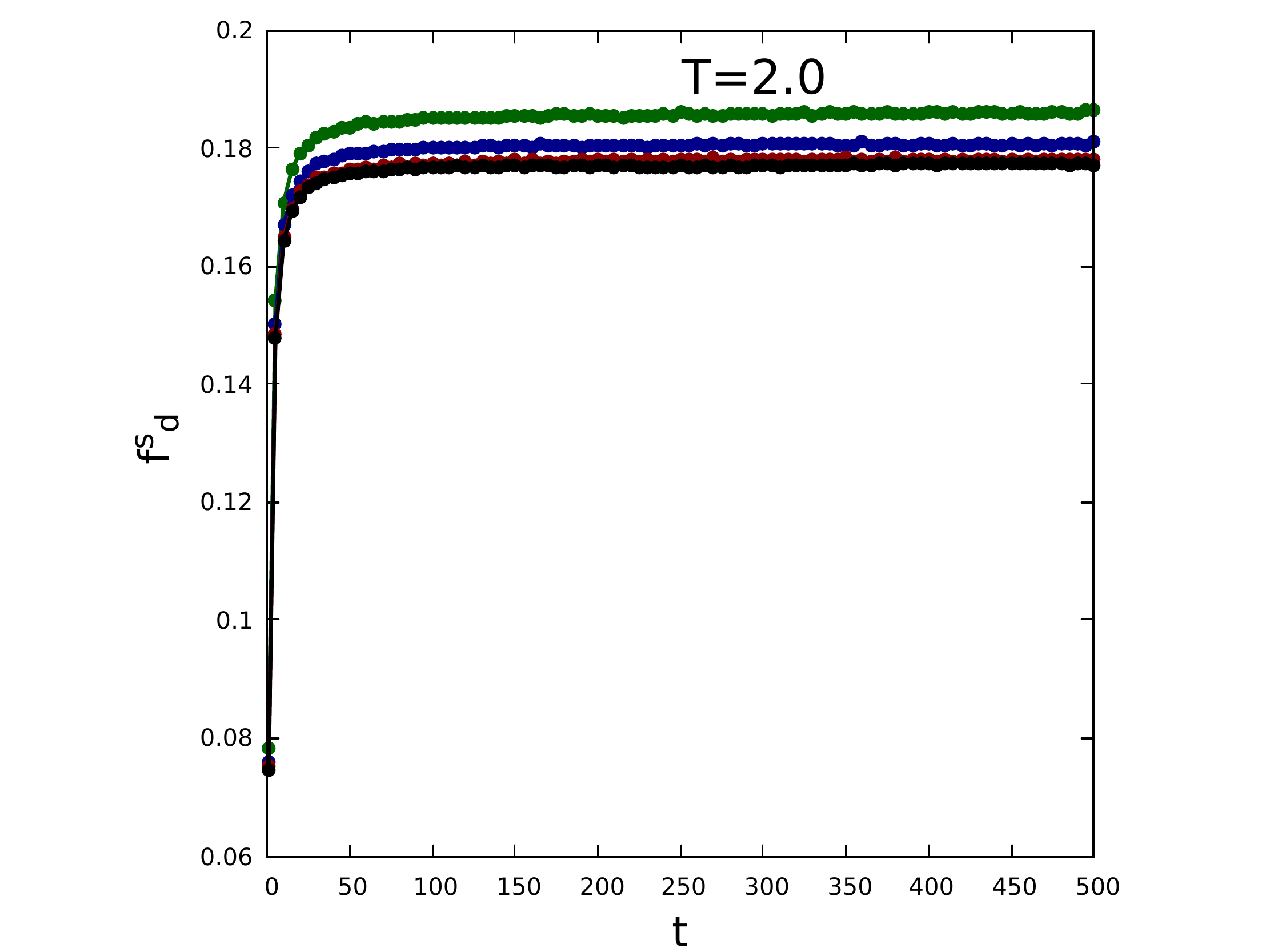}
(b)
\\
\includegraphics[angle=0,width=0.4\textwidth]{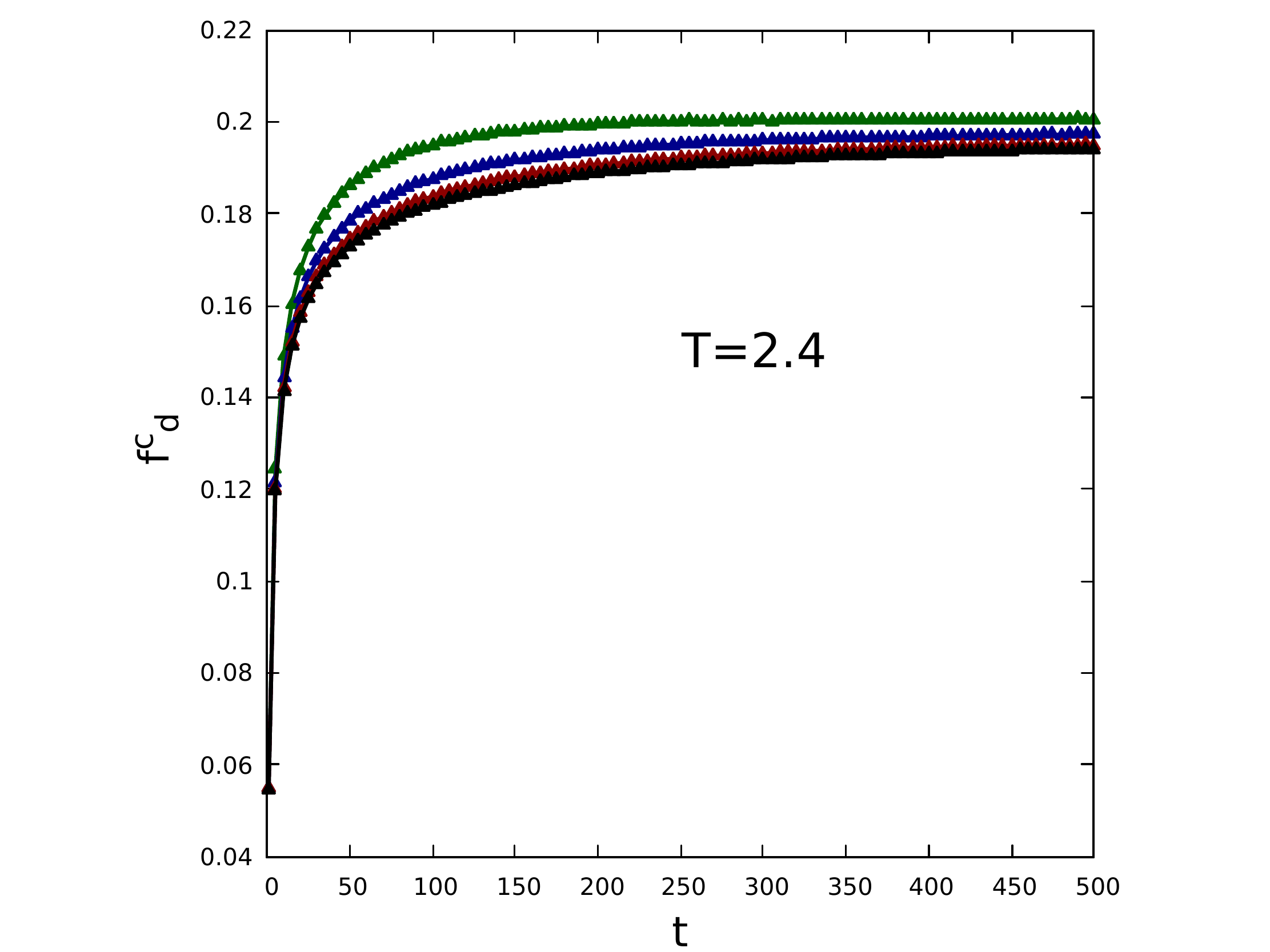}
(c)
\includegraphics[angle=0,width=0.4\textwidth]{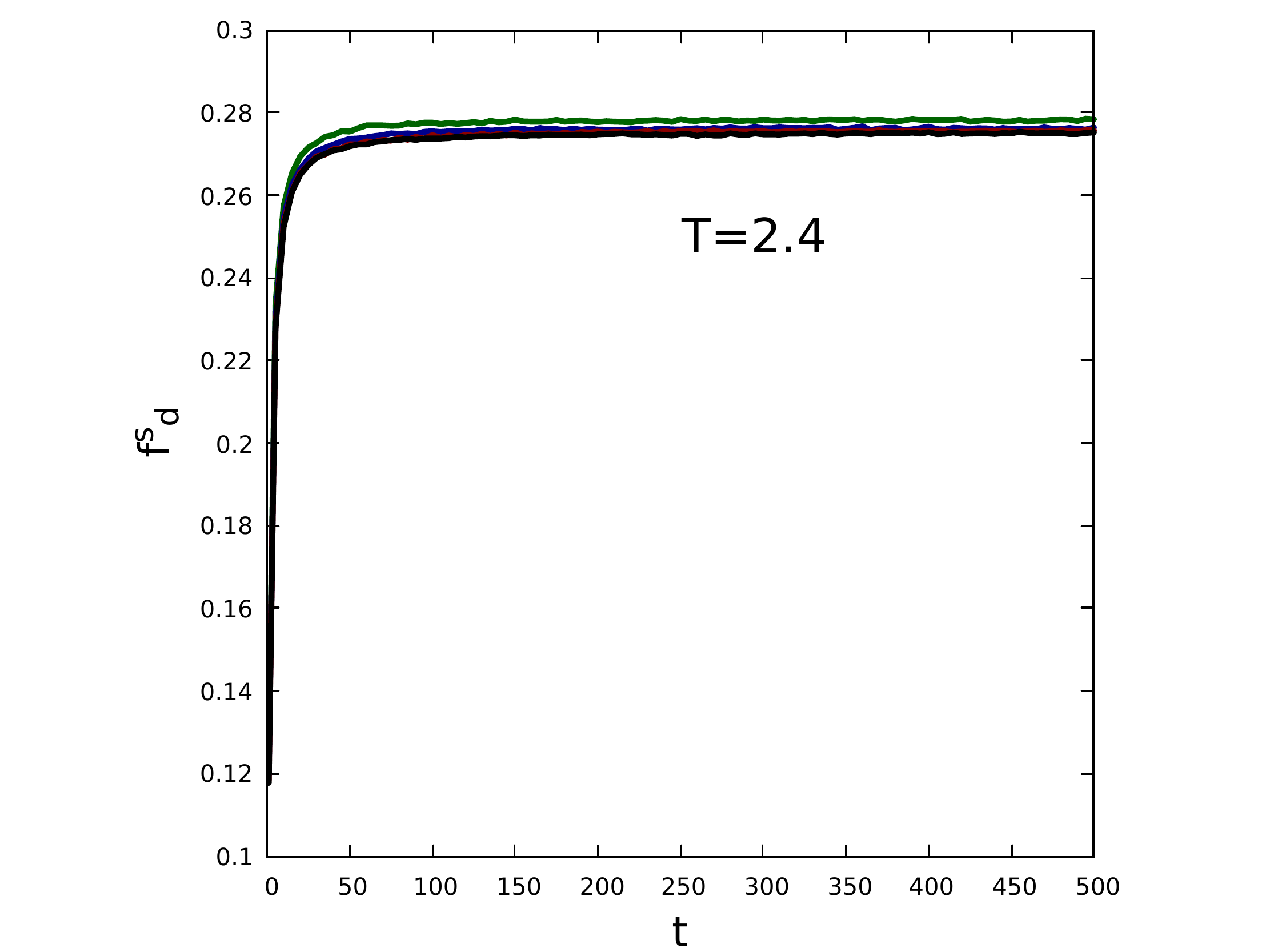}
(d)

\caption{(a) Core spin flip density is plotted against the time(MCSS)  at the temperature $T=2.0$ for different system sizes ($L$); $L=200$(black), $L=100$(red), $L=50$(blue), $L=25$(green), (b) Surface spin flip density is plotted against the time(MCSS) at the temperature $T=2.0$ for different system sizes ($L$); $L=200$ (black), $L=100$(red), $L=50$(blue), $L=25$(green), (c) Core spin flip density is plotted against the time(MCSS)  at the temperature $T=2.4$ for different system sizes
($L$); $L=200$(black), $L=100$(red), $L=50$(blue), $L=25$(green), (d)Surface spin flip density is plotted against the time(MCSS)  at the temperature $T=2.4$ for different system sizes; $L=200$(black), $L=100$(red), $L=50$(blue), $L=25$(green). Here all the results are obtained by using the Glauber dynamics.}

\label{fdg}
\end{center}
\end{figure}

 Since we are simulating for a small size of lattice, we have averaged over 10000-80000 different random samples in each different case. The spin-flip density is measured for  the ferromagnetic phase ($T=2.00$) as well as in the paramagnetic phase ($T=2.4$). We are also interested to do a comparative study between the two 
  kinds of dynamics. The core spin-flip density ($f^c_d$) and surface spin-flip density ($f_d^s$) using Glauber dynamics for temperature $T=2.0$ are reported in Fig-\ref{fdg}(a) and
  Fig-\ref{fdg}(b) respectively.  Here we observed that after a certain period of time the spin flip density became saturated to steady values. The spin flip density for different system sizes was found to saturate at different values. At temperature $T=2.0$, the spin flip density was found to increase as the system size ($L$) decreased. Also, it has been observed that the saturated surface spin flip density is higher than that of the core for any given $L$. What will be the scenario in the paramagnetic phase ? We have performed the simulation in the temperature $T=2.4$. Fig-\ref{fdg}(c) and Fig-\ref{fdg}(d) show the time variations of the core and surface spin flip density respectively at $T=2.4$. It is noticed that in paramagnetic phase, the saturated spin flip density is higher than that observed in the ferromagnetic phase ($T=2.0$). 
Therefore, Glauber kinetic Ising ferromagnet shows the relaxation behaviour in such a way that the effect of thermal excitation is more prominent on the surface than that of core. This is  confirmed by the spin flip density which shows higher values on the surface than that of core.

We have also studied the spin-flip density (both for surface and core) using the
Metropolis dynamical rule. We have observed similar kind of temporal growth of
spin-flip density. However, the saturated spin-flip density is found 
to be higher obviously for Metropolis rule.

\begin{figure}[h!]

\includegraphics[angle=0,width=0.5\textwidth]{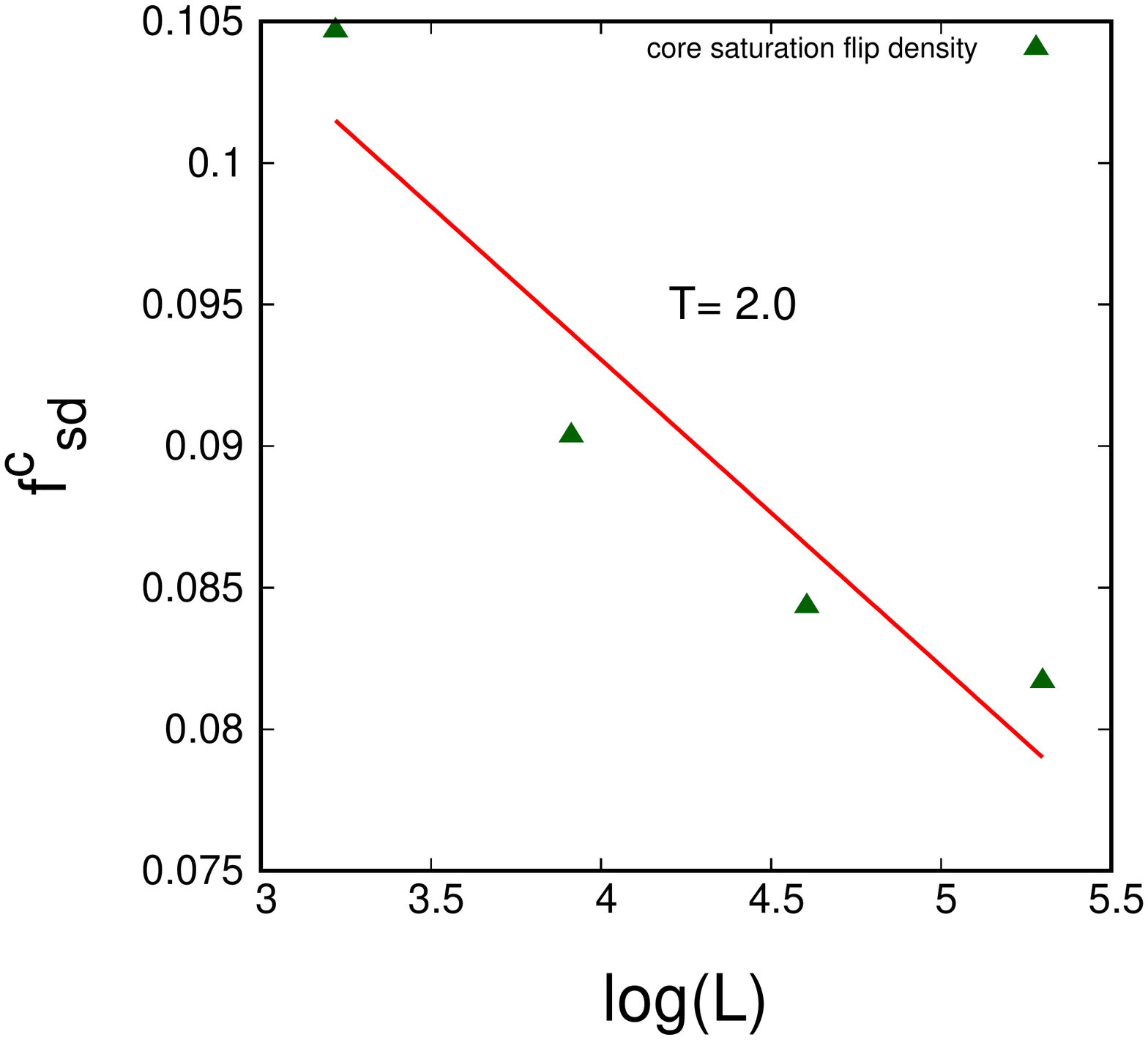}
\includegraphics[angle=0,width=0.5\textwidth]{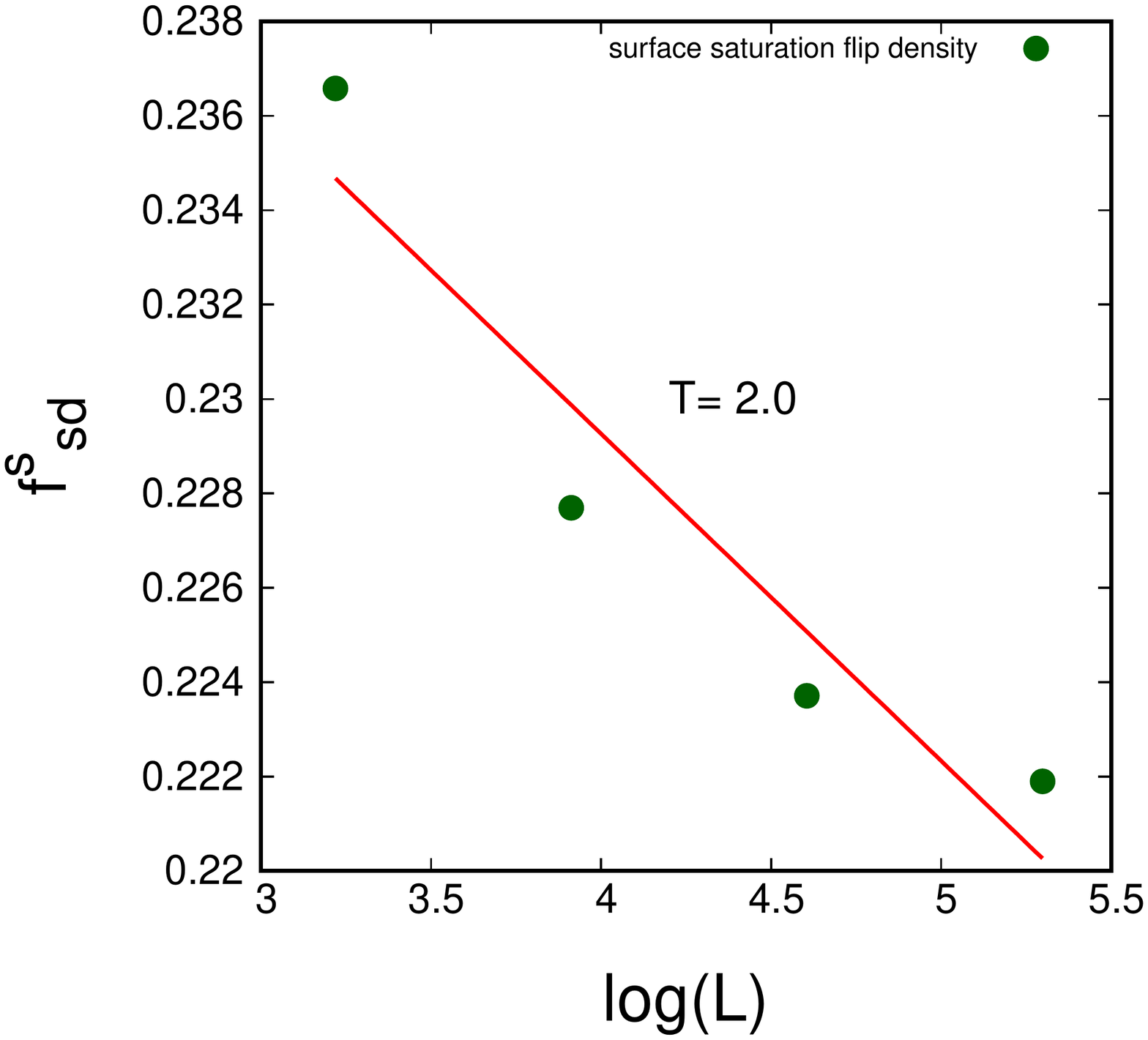}\\

\includegraphics[angle=0,width=0.5\textwidth]{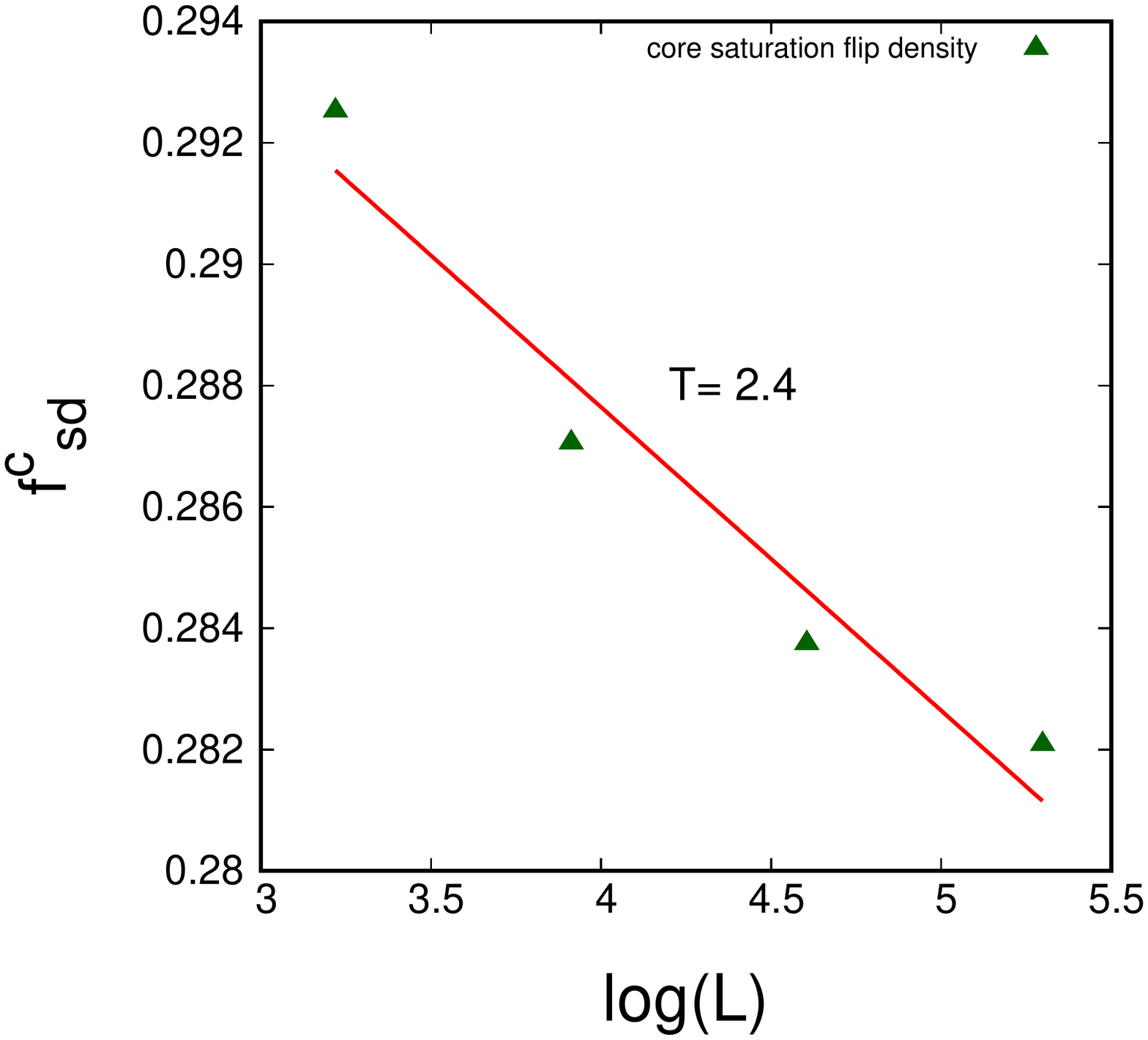}
\includegraphics[angle=0,width=0.5\textwidth]{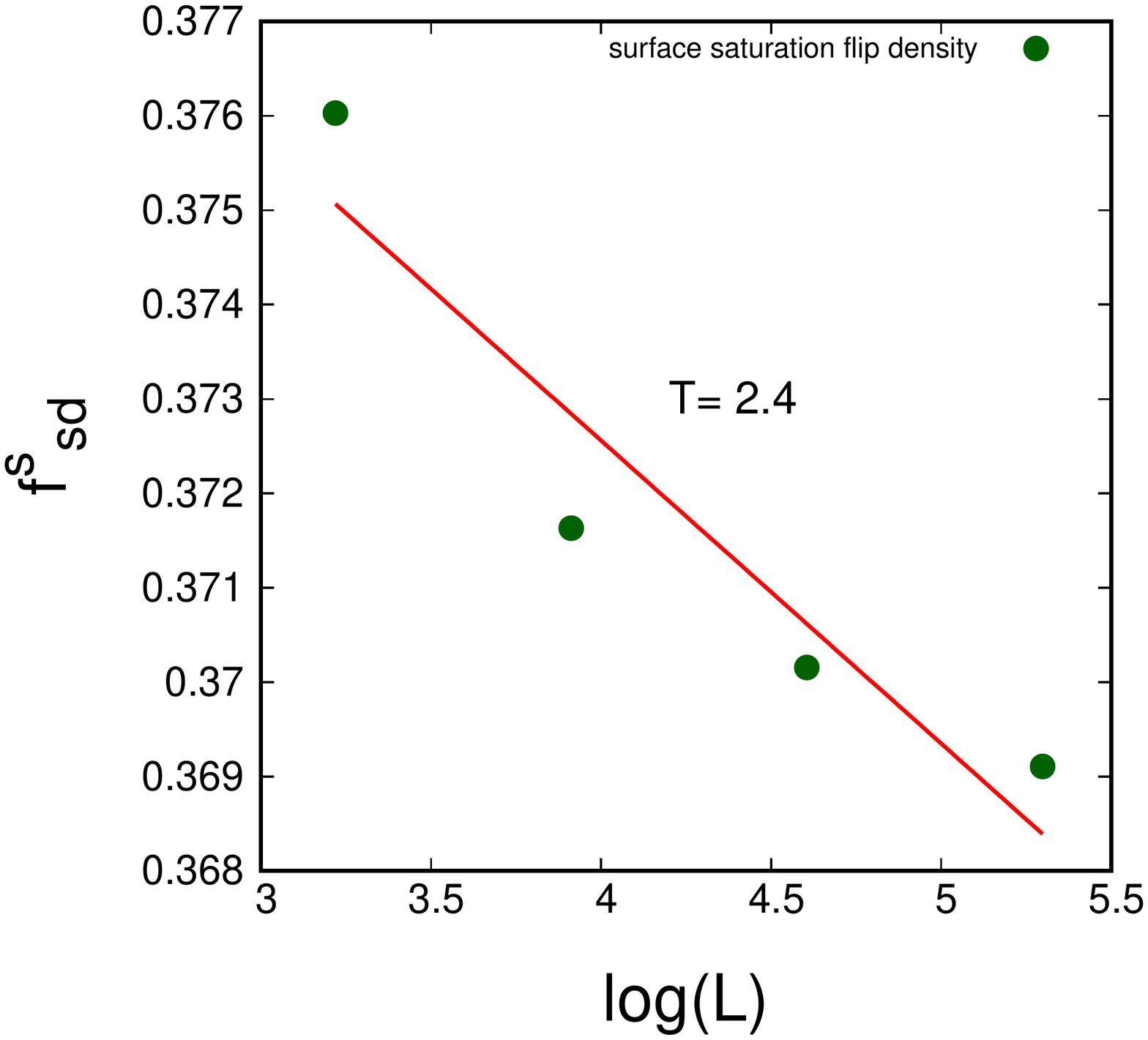}

\caption{Plot of saturation spin flip density  versus $log(L)$  for core
($f^c_{sd}$) and surface ($f^s_{sd}$) for the temperatures $T=2.0$ and $T=2.4$. Data have been obtained by using Metropolis dynamics. The core and surface are labelled as triangles and bullets respectively. Data are fitted to the function $f(L)=a+b~log(L)$. The values of coefficients ($a,b$) and $\chi^2$ are given in Table-\ref{Table3}}.
\label{fdlfitmetro}
\end{figure}

In order to find the saturated spin-flip density, we have averaged (in time) the spin flip density over the last 100 MCSS. The saturated spin-flip  density has been studied with the system size ($L$). The saturated spin flip density (core/surface) has been studied as linear function of $log(L)$.  These are fitted to the function $f(L)=a + b~log(L)$. We have studied both for Glauber and Metropolis dynamics. Each
one is investigated both in ferromagnetic ($T=2.0$) and paramagnetic ($T=2.4$) regions. 
Here, we have shown (in Fig-\ref{fdlfitmetro}) only the results obtained by using the Metropolis dynamics. Similar results are obtained by using Glauber protocol.
 The estimated fitting parameters ($a,b$) along with goodness of fit ($\chi^2$) for each case are shown in
Table-\ref{Table2} and in Table-\ref{Table3} for Glauber and Metropolis dynamics
respectively.

\begin{table}[h!]
\begin{center} {\bf{Table 2}} \\
 \vspace{0.2cm}   
\begin{tabular}{|p{2.0cm}|p{2.0cm}|p{2.5cm}|p{2.5cm}|p{2cm}|p{1cm}|}
  \hline
  Temperature  &  & a & b & ${\chi }^2$  & DoF \\ [1ex]
  \hline 
   \multirow{2}{*}{2.0} & Core & $0.099\pm0.006$& $-0.006\pm0.001$ & 0.000012 &  2 \\ [1ex] \cline{2-6}
   & Surface & $0.198\pm 0.005$ & $-0.004\pm 0.001$ & 0.000005 & 2 \\ [1ex]
   \hline
   \multirow{2}{*}{2.4} & Core & $0.210 \pm 0.002$ & $-0.003 \pm 0.0005 $ & 0.000001 &  2 \\ [1ex] \cline{2-6} 
   & Surface & $0.282\pm 0.001$ & $-0.001\pm 0.0003$ & 0.000007 & 2 \\ [1ex]
   \hline
  \end{tabular}
 \caption{Fitting parameters for spin flip density versus logarithm of system size. Data obtained by using Glauber Dynamics.}
 \label{Table2}
\end{center}
\end{table}

\vskip 0.5 cm

\begin{table}[h!]
\begin{center} {\bf{Table 3}} \\
 \vspace{0.2cm}   
\begin{tabular}{|p{2cm}|p{2cm}|p{2.5cm}|p{2.5cm}|p{2cm}|p{1cm}|}
  \hline
  Temperature  &  & a & b & ${\chi }^2$  & DoF \\ [1ex]
  \hline 
   \multirow{2}{*}{2.0} & Core & $0.136\pm0.012$& $-0.011\pm0.003 $ & 0.000035 &  2 \\ [1ex] \cline{2-6}
   & Surface & $0.257\pm 0.007$ & $-0.007\pm 0.001$ & 0.000013 & 2 \\ [1ex]
   \hline
   \multirow{2}{*}{2.4} & Core & $0.308\pm 0.004$ & $-0.005 \pm 0.0009 $ & 0.000003 &  2 \\ [1ex] \cline{2-6} 
   & Surface & $0.385\pm 0.003$ & $-0.0032\pm 0.0008$ & 0.000003 & 2 \\ [1ex]
   \hline
  \end{tabular}
 \caption{Fitting parameters for spin flip density versus logarithm of  system size. Data obtained by using Metropolis Dynamics.}
 \label{Table3}
\end{center}
\end{table}

\newpage


\noindent {\Large{\bf 4. Concluding remarks:}}

 In this article, we have reported the results of our simulational studies on the relaxation behaviour
 of two-dimensional Ising ferromagnet. We have investigated the role of geometry, boundary conditions, and the dynamical rules on the relaxation behavior of Ising ferromagnet.

 Our primary observations are (i) the relaxation of the magnetisation ($m(t)$) is exponential in nature
 ($m(t) \sim e^{(-t/{\tau})})$. (ii) The relaxation time ($\tau$)  has significant dependence on the aspect ratio ($R$), in the large $R$ limit.  (iii) The relaxation time ($\tau$) shows a power law dependence on the aspect ratio ($R$), like 
 $\tau \sim R^{-s}$. (iv) The exponent ($s$) has been found to be
 linearly dependent ($s(T) = aT+b$) on the temperature ($T$). 
  The above-mentioned observations
 are universal and found in both (Metropolis and Glauber) kinds of dynamics
 and for both (open and periodic) types of boundary conditions. 
 The values of $a$ and $b$ have weak dependence on the dynamical rules and strong 
 dependence on the boundary conditions. The magnitudes of $a$ and $b$ are higher
 for the periodic boundary conditions. The values of $a$ and $b$ are given
 in tabular form in Table-\ref{Table1}. We believe that these observations are new and systematically presented in this manuscript. The faster relaxation was
 observed for open boundary condition for any kind (Metropolis/Glauber) of dynamical rule. 
 Similarly, Metropolis algorithm yields faster relaxation for any 
 kind (open/periodic) of boundary condition.

 We have also studied the temporal growth of instantaneous spin-flip density (already defined in the text). Here, the boundary
 conditions are kept open to investigate the behavior in the bulk and on the surface of the system. However, the behavior of the spin-flip density has been studied both in the ferromagnetic and paramagnetic regions. The behavior of spin-flip density has also been investigated for different (Metropolis 
 and Glauber) dynamical rules. We have observed a logarithmic 
  dependence ($f_d = a + b~log(L))$ of saturated spin-flip density with 
  system size ($L$), both for core and surface spin flip density. The 
  logarithmic dependence has been observed both in ferromagnetic and paramagnetic regimes. The values of $a$ and $b$ (for such logarithmic fitting) and the values of 
  $\chi^2$ are given in Table-\ref{Table2} and in Table-\ref{Table3} for Glauber
  and Metropolis dynamics respectively.

\vskip 0.5cm

\noindent {\Large {\bf Acknowledgements:}} IT acknowledges the UGC-JRF fellowship, the Government of India, for financial support. OM acknowledges the MANF, UGC, Government of India for financial support.

\vskip 0.5cm
\noindent{\Large{\bf {\underbar {Declarations:}}}}

\noindent {\bf Data availability statement:} The Data may be available on request to Ishita Tikader.

\vskip 0.1cm

\noindent {\bf Conflict of interest:} The authors have no financial or proprietary interests in any material discussed in this article.

\vskip 0.1cm

\noindent {\bf Funding:} No funding was received particularly for this research.

\vskip 0.1cm
\noindent {\bf Authors' contributions:} Ishita Tikader-Developed the code,
obtained the data, prepared the figures, wrote the manuscript, Olivia Mallick-Developed the code,
obtained the data, prepared the figures, wrote the manuscript, Muktish Acharyya- Conceptualized the problem, Analysed the result, wrote the manuscript.

\newpage

\newpage

\newpage

\newpage


\end{document}